# Simultaneously monitoring Ga adsorption and desorption kinetics on GaN(0001) using four in situ techniques

Huaide Zhang,[a] Philipp John, Jingxuan Kang, Lutz Geelhaar, Yongjin Cho, and Oliver Brandt
*Paul-Drude-Institut für Festkörperelektronik, Leibniz-Institut im Forschungsverbund Berlin e.V., Hausvogteiplatz 5-7, 10117 Berlin, Germany*

We present a systematic investigation of Ga adsorption and desorption kinetics on the wurtzite GaN(0001) surface using four *in situ* techniques operated simultaneously: reflection high-energy electron diffraction, laser reflectometry, line-of-sight quadrupole mass spectrometry, and optical pyrometry. Flux- and temperature-dependent experiments are performed for Ga coverages ranging from the submonolayer to the droplet regime. Despite their distinct transient responses, the signals from all four techniques and their trends with surface coverage are quantitatively reproduced by a unified kinetic model of Ga adsorption, diffusion, and desorption. An Arrhenius analysis of the Ga adlayer desorption yields an activation energy of $(2.87 \pm 0.04)$ eV.

## I. INTRODUCTION

Complex semiconductor heterostructures required for modern electronic and optoelectronic devices are commonly synthesized by either metal–organic chemical vapor deposition (MOCVD)[1,2] or molecular beam epitaxy (MBE).[3–5] While MOCVD is the dominant technique for large-scale production owing to its high throughput and scalability, MBE provides a uniquely suited platform for fundamental studies of growth mechanisms and the development of novel materials. Its ultra-high vacuum environment facilitates the use of a wide range of *in situ* diagnostic techniques, which have played a central role in establishing the microscopic understanding of epitaxial growth for many III–V semiconductors.

For conventional III–V compounds such as arsenides and antimonides, the surface during growth typically exhibits well-defined reconstructions that can be monitored by reflection high-energy electron diffraction (RHEED).[6,7] Under steady-state conditions, these reconstructions correspond to iso-contours of surface stoichiometry and thus provide a direct means to control growth conditions. The situation is fundamentally different for the group-III nitrides AlN, GaN, and InN in plasma-assisted MBE. While several reconstructions of, for example, the Ga-terminated GaN(0001) surface have been observed at low temperatures,[8–10] they are absent at growth temperature, where the Ga adatoms form a highly mobile, essentially liquid-like adlayer.[11] Consequently, the steady-state surface during growth of GaN does not provide the structural fingerprints commonly used to control the synthesis of other III–V compounds.

In the absence of stable surface reconstructions, valuable information about the surface stoichiometry and kinetics can instead be obtained from the *transient* response of the surface upon temporal changes of the surface stoichiometry. A variety of *in situ* techniques have therefore been employed to study the adsorption and desorption dynamics of Ga on GaN surfaces, including RHEED,[12–15] spectroscopic ellipsometry,[16,17] laser-based optical probes,[18,19] line-of-sight quadrupole mass spectrometry,[20,21] and optical pyrometry.[22,23] Combined with theoretical work, largely based on density functional theory, these studies have led to a qualitative understanding of the surface kinetics during plasma assisted MBE growth.[4,5,24,25] At typical growth temperatures, Ga first forms a metallic adlayer that grows approximately layer-by-layer. Once a critical coverage is exceeded, excess Ga nucleates into nanoscopic clusters that evolve via surface diffusion and coalesce into microscopic droplets.[26] At the same time, the desorption of the adlayer can be significantly delayed by replenishment from excess Ga. Beyond the Ga-polar (0001) face, stable metallic Ga adlayers have also been identified on other surfaces, including GaN($000\bar{1}$),[26,27] GaN($10\bar{1}0$),[28] and GaN($11\bar{2}0$).[29]

The transient signals recorded with the above techniques have also been used to extract fundamental kinetic parameters, most notably the activation energy for Ga adlayer desorption. However, the reported values vary widely in the literature, ranging from approximately 2.4 to 4.8 eV.[12–14,26,30–34] This spread reflects not only the complexity of the underlying surface processes, but also the fact that the different diagnostic techniques probe distinct physical quantities and yield transients with different lineshapes. A reliable determination of the desorption energy therefore requires a quantitative framework that relates the transient signals to the surface coverages and enables the associated kinetic processes to be disentangled. Despite the extensive use of these methods, a systematic comparison of their response and a unified interpretation of the corresponding signals remain lacking.

In this work, we simultaneously monitor Ga adsorption and desorption on the GaN(0001) surface in a plasma-assisted MBE environment by four *in situ* techniques: RHEED, laser reflectometry (LR), line-of-sight quadrupole mass spectrometry (QMS), and optical pyrometry (OP). The temporal evolution of the signals from these techniques is analyzed for two experimental series in which either the Ga flux or the substrate temperature is systematically varied. We show that, although the individual techniques exhibit markedly different signal responses, their behavior can be consistently reproduced by a unified kinetic model of Ga adsorption and desorption. This physically motivated framework allows us to quantitatively correlate the measurements and to extract key parameters of the Ga adlayer kinetics with improved accuracy.

[a] Electronic mail: zhang@pdi-berlin.de

## II. EXPERIMENTS

This study was carried out in a custom-designed three-chamber plasma-assisted MBE system equipped with solid sources for the group-III metals, a plasma cell for providing active $N^*$, and four in-situ diagnostic techniques, namely, RHEED, LR, QMS and OP as illustrated in Fig. 1. The base pressure of the system is well below $10^{-10}$ Torr. During our experiments, for which the substrate and Ga effusion cell were ramped up to 550 – 720 °C and 800 – 920 °C, respectively, the chamber pressure was maintained below $10^{-8}$ Torr. The absolute impinging Ga flux was determined by the thickness of GaN layers grown under slightly N-rich conditions at a temperature where Ga desorption can be neglected. The thickness of these calibration samples was determined by scanning electron microscopy from its cleaved cross-section with an uncertainty of 5%. While these data points yield a value for the absolute flux, more precise values for the change of flux with temperature were obtained by measuring the beam equivalent pressure over a wide temperature range using a retractable ion gauge. The overall uncertainty for the relative impinging flux is thus better than 2%.

As a substrate for our experiments, we used a commercially available, unintentionally doped GaN/$Al_2O_3$ template fabricated by MOCVD. The GaN layer had a thickness of approximately 4 µm, and the template dimensions were $2 \times 2\,\mathrm{cm}^2$. A 1 µm-thick Ti layer was coated on the backside of the substrate to facilitate radiative heating. The surface of the template exhibited a smooth morphology with a root-mean-square (RMS) roughness of 0.16 nm over a $5 \times 5\,\mu\mathrm{m}^2$ area (see Fig. S1a in the Supplementary Material), as measured by atomic force microscopy (AFM). Prior to the adsorption–desorption experiments, a 100 nm-thick GaN buffer layer was grown under Ga-stable conditions by plasma-assisted MBE in the same reactor. The surface remained smooth with an RMS value of 0.37 nm (see Fig. S1b in the Supplementary Material). The static $1 \times 1$ RHEED patterns obtained from both the GaN template and the buffer layer were bright and streaky, accompanied by well-defined Kikuchi lines, reflecting a clean, smooth GaN surface with Ga polarity (see Fig. S2a in the Supplementary Material). During growth of the buffer layer, the $1 \times 1$ pattern persists but becomes markedly dimmer due to the the formation of a complete Ga bilayer (see Fig. S2b in the Supplementary Material).

Two sets of experiments were conducted to investigate the effects of Ga coverage, ranging from submonolayer coverage to droplet formation: a flux-dependent series and a temperature-dependent series. In the flux-dependent experiments, the substrate temperature was maintained at 708 °C, while the impinging Ga fluxes were 0.04, 0.38, and 0.64 monolayer (ML)/s, respectively [1 ML corresponds to $1.135 \times 10^{15}$ Ga adatoms $\mathrm{cm}^{-2}$ on GaN(0001)]. In the temperature-dependent series, the Ga flux was held constant at 0.38 ML/s, while the substrate temperature was varied from 660 to 720 °C. In each case, the Ga shutter was opened for 60 s.

During this Ga supply, the surface was probed by the four *in situ* techniques simultaneously to detect the change of surface stoichiometry upon the adsorption of Ga and its subsequent desorption. The RHEED pattern was monitored in $[11\bar{2}0]$ azimuth using an acceleration voltage of 20 kV and a comparatively steep incidence angle of about 4°. The intensity of the fundamental 00 diffraction streak including the specular spot was recorded using a charge-coupled device (CCD) camera controlled by a digital RHEED analysis system.[35] To minimize the impact of electromagnetic interferences, the growth chamber is encased in an active compensation cage. The custom-designed LR system integrated into our MBE setup consists of a 650 nm laser diode module with an optical power of 1 mW, a Si photodiode with a polarization layout, and adjustable optical mounts. The reflected laser intensity was recorded and analyzed using in-house software specifically developed for this purpose. The line-of-sight mass spectrometer is mounted in an effusion cell port with an angle of 37.3° relative to the substrate normal. An aperture between the QMS ionizer and the substrate restricts the line-of-sight acceptance angle to ensure that the ionizer accepts only Ga atoms desorbed from the substrate area. The QMS response to the $^{69}$Ga signal was calibrated in units of an equivalent flux. The optical pyrometer installed in our system detects black-body radiation with a maximum sensitivity at 900 nm, enabling real-time temperature monitoring of the sample. Since the GaN template is transparent at this wavelength, the pyrometer monitors the temperature of the Ti coating at the backside of the wafer. An emissivity of 0.55 was assumed for Ti.[36]

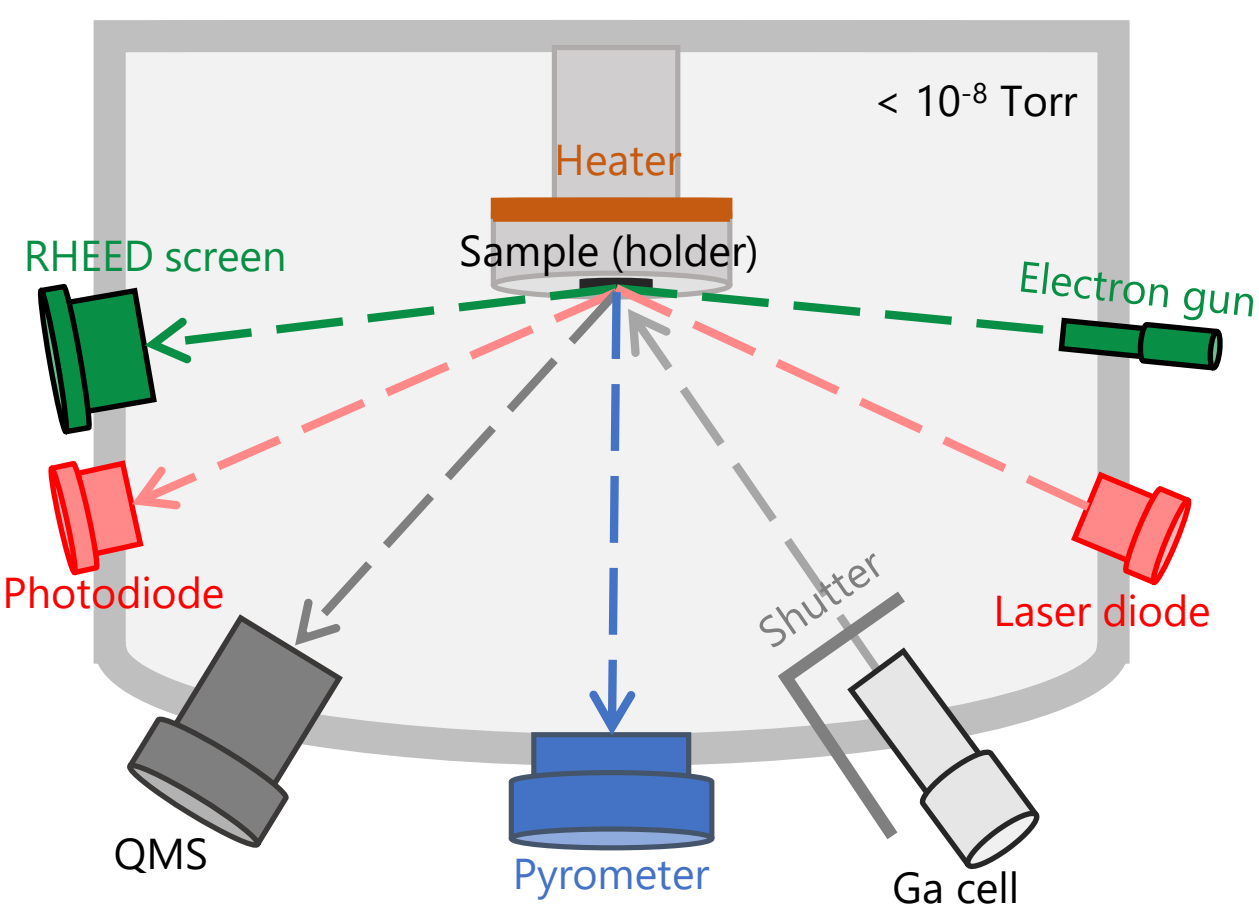

FIG. 1. **Schematic of our experimental setup.** Depicted are the MBE growth chamber with the sample holder, the shuttered Ga effusion cell, and the four *in situ* techniques it is equipped with, namely, RHEED, LR, QMS, and OP.

## III. RESULTS AND DISCUSSION

Figure 2 compares the experimental transient signals detected by the four *in situ* techniques upon supplying a Ga dose to the surface with corresponding simulations. The underlying model has been developed in previous work to describe the Ga adsorption-diffusion-desorption kinetics on GaN($10\bar{1}0$) surfaces.[28] The model accounts for adsorption and desorption of Ga as well as for the formation of excess Ga interacting with the Ga adlayer. In units normalized to the maximum Ga adlayer coverage $\theta_m$ (2.4 ML),[14] the model is described by the following system of coupled non-linear first-order differential equations:

$$\theta'(t) = \phi_{\text{Ga}}(t)\,[1-\theta(t)] + \delta\,[1-\theta(t)]\,\nu(t) - \gamma\theta(t), \quad \text{(1a)}$$

$$\nu'(t) = \phi_{\text{Ga}}(t)\theta(t) - \delta\,[1-\theta(t)]\,\nu(t) - \kappa(t)\nu(t). \quad \text{(1b)}$$

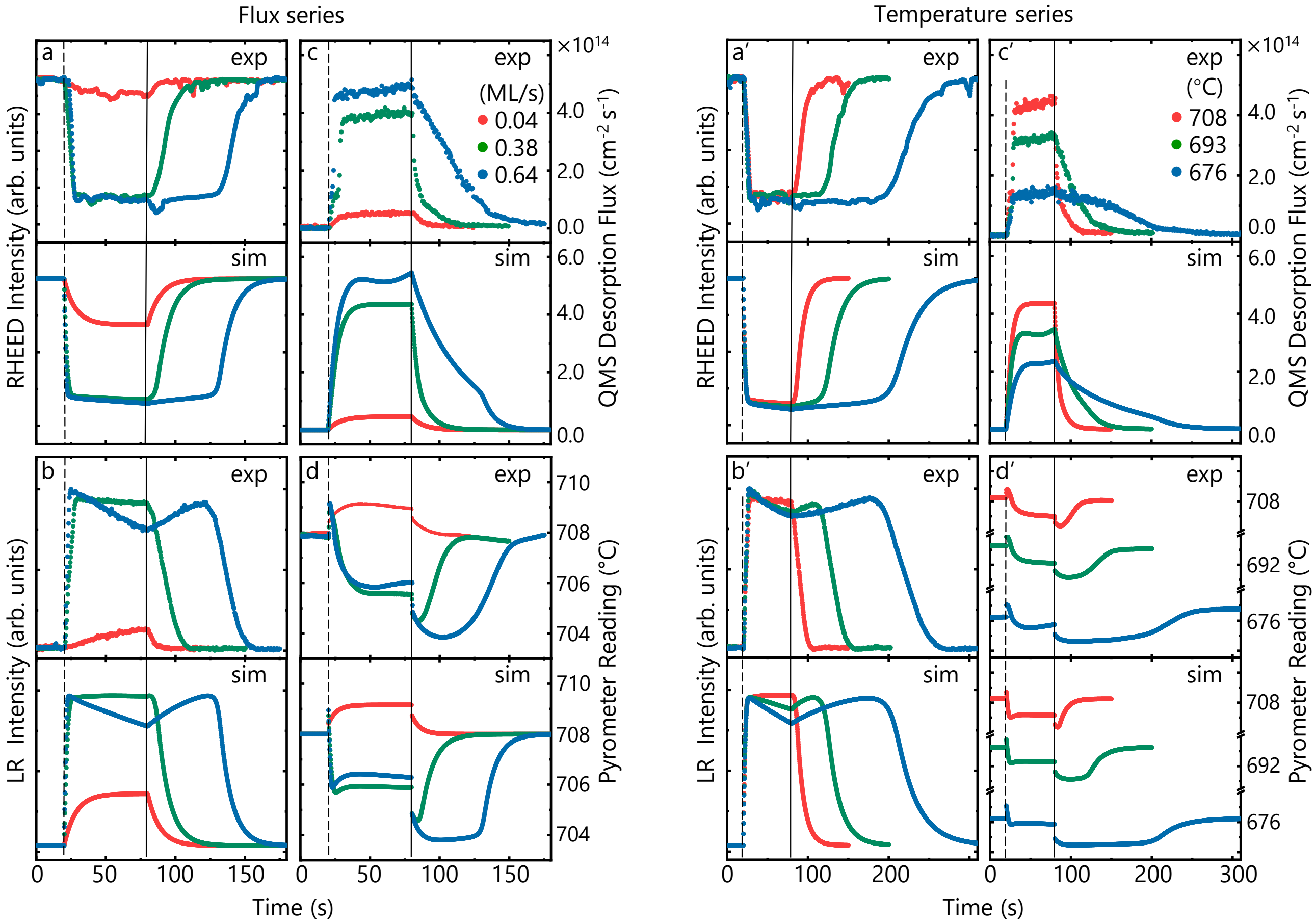

FIG. 2. **Transient signals detected simultaneously by four *in situ* techniques during the adsorption and desorption of Ga on GaN(0001).** Experiments conducted with constant temperature of 708 °C and different Ga flux (constant flux of 0.38 ML/s and different temperatures) are depicted on the left (right). Panels **a, a'** show the transient RHEED intensity, **b, b'** the reflectivity, **c, c'** the Ga desorption flux, and **d, d'** the apparent change in substrate temperature. The vertical dashed and solid lines indicate the times $t$ for opening ($t = 20$ s) and closing ($t = 80$ s) the Ga shutter, respectively. The labels "exp" and "sim" denote the experimental and simulated transients in the respective top and bottom panels.

Here, $\theta$ and $\nu$ denote the Ga adlayer and the excess Ga coverage, respectively, $\phi_{\text{Ga}}$ the Ga delivery rate, $\delta$ the diffusion rate of excess Ga adatoms impinging onto $\theta$, $\gamma$ and $\kappa$ the desorption rate constant of the adlayer and excess Ga, respectively. The first term of each equation accounts for adsorption of Ga, building up the Ga adlayer $\theta$ and subsequently excess Ga $\nu$. In much the same way, the second terms account for diffusion of excess Ga impinging onto the Ga adlayer. The third terms describe the desorption of $\theta$ and $\nu$. The time-dependent rate constant $\kappa(t)$ accounts for the ripening of excess Ga clusters into droplets. Since Ga desorption occurs from the surface of these three-dimensional agglomerates, the desorption rate decreases proportional to their size. For simplicity, we assume that the rate constant decreases linearly with time from an initial value of $\kappa_0$ once Ga deposition is initiated.

In the following, we will discuss the results presented in Fig. 2 for each technique separately. We start with Figs. 2a and 2a', which present the changes in RHEED intensity reflecting the ad- and desorption processes of Ga. The RHEED intensity rapidly drops once Ga deposition is initiated, and recovers either immediately or with a delay once the Ga supply is stopped. This behavior is well understood from previous studies:[12,13,37–39]the drop originates from Ga adsorption and the formation of the adlayer, and the recovery from the desorption of the adlayer. The delay is a consequence of excess Ga acting as a reservoir, replenishing the adlayer and thus maintaining complete coverage. The slope of the intensity recovery increases with temperature in Fig. 2a', directly reflecting the higher desorption rate of the Ga adlayer.

For a quantitative interpretation of these results, we have to establish a relation between the surface coverages calculated by means of Eqs. (1a) and (1b) and the RHEED intensity. Since the Ga adlayer at growth temperature does not condense into a reconstruction, but rather represents a liquid, it manifests itself in an attenuation of the RHEED intensity. Droplets cause a further decrease in intensity, primarily due to shadowing. We thus take the well-established relation between surface coverage and intensity in Auger attenuation experiments,[40] which reads

$$I(t) = \exp\left[-\theta(t) - a_1\nu(t)\right], \tag{2}$$

where both $\theta$ and $\nu$ are assumed to attenuate the intensity exponentially, with $a_1$ being a scaling coefficient depending on the geometry of the experiment. In the present work, $a_1$ is small due to the comparatively steep incidence angle, but still improves the agreement of the simulation with the experimental transients for the highest coverage. Excellent agreement with the experimental transients is obtained by adjusting $\delta$ and $\kappa(t)$, which govern the delay preceding adlayer desorption, and $\gamma$, which controls the slope during the desorption stage. The resulting temporal evolution of $\theta$ and $\nu$, together with the values of $\gamma$, $\kappa_0$ and $\delta$ used for these and all subsequent simulations shown in Fig. 2, are provided in Fig. S3 and Table S1 of the Supplementary Material, respectively.

Figures 2b and 2b' show the transient changes in laser reflectivity upon Ga adsorption and desorption. The reflectivity rapidly increases upon Ga adlayer formation and decreases when excess Ga forms. Qualitatively, the well-ordered Ga adlayer acts as a mirror, effectively reflecting the incident laser beam, whereas the droplets cause diffuse scattering, hence leading to a reduction in reflectivity. For a quantitative modeling, we assume that the reflectivity is related to the surface coverage via the simple expression

$$R(t) = r_0\left[1 - \theta(t)\right] + r_1\theta(t)\exp\left[-a_2\nu(t)\right], \tag{3}$$

where $r_0$ and $r_1$ represent the reflectivity of the bare GaN surface and of the surface covered with an Ga adlayer, respectively, while $a_2$ represents the attenuation coefficient associated with light scattering by agglomerated excess Ga. Keeping the values of $\delta$, $\kappa(t)$, and $\gamma$ the same as assumed for the simulations of the RHEED intensity transients in Figs. 2a and 2a', we obtain a satisfactory match between the simulated and experimental reflectivity transients by adjusting $r_0$, $r_1$, and $a_2$.

The behavior of the Ga desorption flux measured by QMS and shown in Figs. 2c and 2c' is straightforward to understand. As expected, the desorbing flux generally increases with either higher impinging Ga fluxes (Fig. 2c) or substrate temperatures (Fig. 2c'). For high Ga coverages, the recovery is delayed for the same reason as in RHEED and LR, namely, excess Ga feeding the adlayer and thus preventing its desorption. The total desorption rate $\phi$, which includes contributions from both the adlayer and excess Ga, is simply given by:

$$\phi(t) = \gamma\theta(t) + \kappa(t)\nu(t). \tag{4}$$

Once again, we use the same values for the parameters $\delta$, $\kappa(t)$, and $\gamma$ as assumed for the simulations shown in Figs. 2a–2b and 2a'–2b'. The absolute values for the desorbing Ga flux are well reproduced, as well as the delay observed for the highest coverages.

Figures 2d and 2d' present the pyrometer temperature readings. Of all techniques studied in this work, the pyrometer signal exhibits the most complex lineshape and defies an intuitive understanding. The signal even exhibits a change of sign: while we observe an *increase* in temperature for the lowest flux in Fig. 2d, the temperature *decreases* after an initial, rapid increase for the higher coverages. In these experiments, we furthermore observe a second decrease of intensity after closing the Ga shutter, which is presumably related to the presence of excess Ga on the surface.

We have identified three different effects that contribute to this coverage-dependent lineshape. First, reflected radiation from the hot effusion cell (≈900 °C) causes an instantaneous increase (decrease) of the apparent substrate temperature of about 0.5°C upon opening (closing) the Ga shutter.[23] Second, the same radiation causes an actual (radiative) heating of the substrate, which is a slower effect, and dominates for the transient with the lowest coverage.[41] Third, the metallic adlayer changes the actual emissivity of the sample, while the temperature determined by the pyrometer software is still based on the emissivity of Ti.[42]

Combining these three effects into one expression,[41–44] we arrive at a pyrometer reading

$$T_P(t) = \Delta T_r(t) + T_2(t)\left(1 + a_4 \ln\frac{\varepsilon(t)}{\varepsilon_0}\right), \tag{5}$$

where $\Delta T_r(t)$ is the instantaneous change due to reflected radiation. $T_2(t)$ is the actual surface temperature, which can be obtained from Newton's law of cooling:

$$T_2'(t) = -\alpha\left[T_2(t) - T_1(t)\right] \tag{6}$$

with the heat transfer coefficient $\alpha$, and

$$T_1(t) = T_0 + \Delta T_h(t). \tag{7}$$

where $T_0$ [$T_1(t)$] is the substrate temperature without (with) the radiative heating from the Ga cell, and $\Delta T_h(t)$ represents this heating. Finally, the emissivity $\varepsilon(t)$ depends on coverage according to

$$\varepsilon(t) = \varepsilon_0[1 - \theta(t)] + \varepsilon_1\theta(t)\exp[-a_3\nu(t)], \tag{8}$$

where $\varepsilon_0$ is the emissivity of Ti, and $\varepsilon_1$ the effective emissivity of the wafer with full Ga adlayer coverage, and $a_3$ the attenuation coefficient due to excess Ga. The logarithmic dependence on $\varepsilon(t)$ with a prefactor $a_4$ accounts for the error arising from assuming an incorrect emissivity.[42]

The simulations shown in Figs. 2d and 2d' are performed using the same values for $\delta$, $\kappa(t)$, and $\gamma$ as for all other simulations in Fig. 2 and are seen to be in satisfactory agreement with the experimental data. In particular, they reproduce the change of sign with increasing coverage, as well as the additional temperature decrease and the delayed recovery for high coverages.

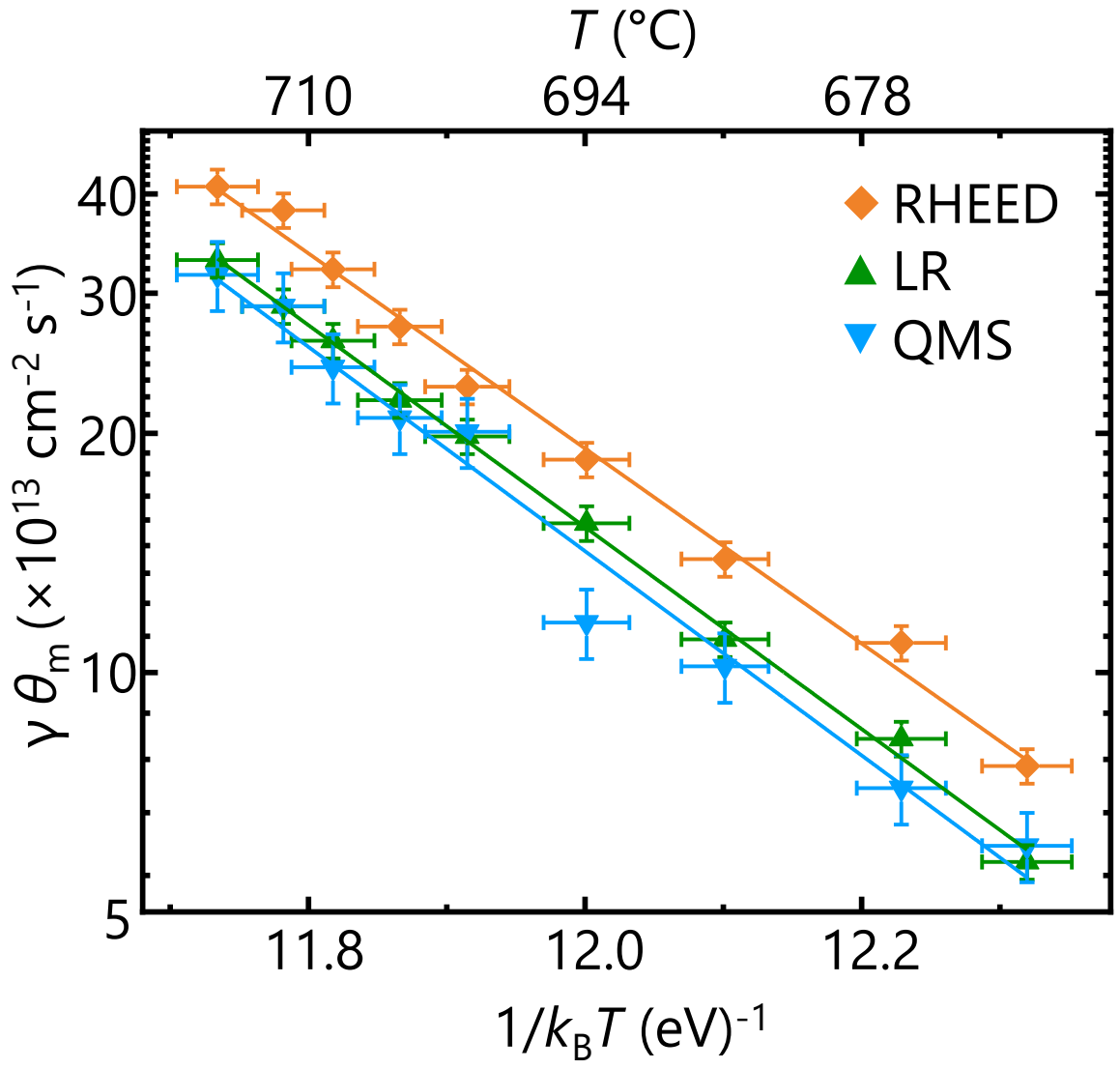

FIG. 3. **Arrhenius representation of Ga adlayer desorption.** The plot shows the temperature dependence of the maximum Ga adlayer desorption flux $\gamma\theta_m \propto \exp(-\frac{E_A}{k_B T})$. Exponential fits (solid lines) to the RHEED (diamonds), LR (upward triangles), and QMS (downward triangles) data yield activation energies $E_A$ of $(2.82 \pm 0.09)$, $(2.87 \pm 0.05)$, and $(2.91 \pm 0.16)$ eV, respectively. Combining these independent determinations yields a weighted mean of $(2.87 \pm 0.04)$ eV.

Having discussed the results obtained from all individual techniques and their simulations, we are now in a position for a critical comparison and a brief discussion of their benefits and drawbacks in the special context of Ga adsorption and desorption kinetics. Most importantly, the transients shown in Fig. 2 clearly demonstrate that all of them are capable of detecting the formation of the Ga adlayer, and are also sensitive to the accumulation of excess Ga interacting with the adlayer. All of these techniques are thus suitable for either fundamental investigations of the Ga adsorption-desorption kinetics, or for the routine calibration of the substrate temperature by determining the adlayer desorption flux. The latter provides a convenient replacement of the RHEED phase diagram used for conventional III-V compounds as a system-independent method to probe the surface stoichiometry.[6,7]

From a practical point of view, RHEED and QMS require a high vacuum chamber, while LR and OP also work at pressures more commonly found in a CVD environment or in a thermal laser epitaxy (TLE) chamber.[45] The instrumental costs for RHEED and QMS are also not insignificant, while a functional LR setup is much more affordable, and can be easily designed and built in-house. At the same time, all of these techniques are primarily used for purposes *other* than studying surface stoichiometry. In particular, RHEED provides such a wealth of information about the state of the growth front that we consider it indispensable for MBE growth regardless of the material system.

From a theoretical point of view, our model reproduces the trends observed by four physically distinct techniques using a simple coupled rate-equation framework combined with analytical expressions that link the adlayer and excess coverages to the corresponding experimental observables. A clear limitation of this approach is the approximate treatment of diffusion and ripening dynamics of excess Ga. Depending on the dominant ripening mechanism, a more rigorous description would require solving the equations governing Ostwald[46] or Smoluchowski[47] ripening in two dimensions, probably most efficiently using Monte Carlo techniques.[48] While such calculations would provide valuable insights into the underlying ripening dynamics, their substantial computational cost would preclude the quasi-real-time simulation of experimental data enabled by the present phenomenological model.

As a byproduct of our simulations, we obtain reliable values of a quantity of particular interest, namely, the adlayer desorption flux and its activation energy $E_A$. Among the four techniques considered, RHEED and LR exhibit the closest agreement between experiment and simulation during the adlayer desorption stage and are therefore expected to provide the most reliable values for $E_A$. For the transients obtained by QMS and OP, disentangling adlayer desorption from the dynamics of the excess Ga coverage is less straightforward. This partial superposition of processes is particularly pronounced for OP, which we therefore consider inherently less suitable for an accurate determination of the adlayer desorption flux.

Figure 3 shows an Arrhenius representation of the maximum adlayer desorption flux (for $\theta = \theta_m$) derived from simulations reproducing the entire series of temperature-dependent RHEED, LR, and QMS transients (cf. Fig. 2a'–c' for three representative transients and their simulations). While the slopes of the corresponding data sets are nearly identical, their absolute values differ systematically. We attribute the lower desorption fluxes obtained from LR and QMS to the larger surface areas probed by these techniques. Since the sample temperature decreases toward the sample edge, averaging over a larger area results in an effectively lower mean temperature. Correcting for this offset of approximately 7 K, exponential fits to the RHEED, LR, and QMS data yield activation energies of $E_A = (2.82 \pm 0.09)$, $(2.87 \pm 0.05)$, and $(2.91 \pm 0.16)$ eV, respectively. The weighted mean of these values is $(2.87 \pm 0.04)$ eV. This result is in close agreement with the values of approximately 2.8 eV reported in several previous studies.[13,30,32,33,49,50]

We emphasize that the present work focuses on comparing the different diagnostic techniques and their ability to detect Ga adsorption and desorption; the experiments were therefore not designed for a precision determination of $E_A$. Nevertheless, the quantitative reproduction of the full transient lineshape provides a more robust and internally consistent determination of $E_A$ than approaches based on characteristic time intervals.[28] In particular, it establishes the relation between the measured signals and the surface coverages and allows the individual kinetic contributions to be resolved.

## IV. CONCLUSION

We have studied the adsorption and desorption kinetics of Ga adatoms on the GaN(0001) surface by simultaneously monitoring the surface with RHEED, LR, QMS, and OP while systematically varying the impinging Ga flux and substrate temperature. The resulting data are well reproduced by a simple rate-equation model. This agreement demonstrates that the model, together with the expressions developed here that link it to the experimental observables, provides a reliable and robust description of Ga adatom behavior across all diagnostic methods. Beyond the specific case of PAMBE growth of GaN, the insights obtained here are relevant for other growth approaches for III-nitrides, including reactive MBE and TLE, where *in situ* diagnostics likewise provide essential information about the surface processes governing epitaxial growth.

## SUPPLEMENTARY MATERIAL

See the supplementary material for AFM topographs (Fig. S1) and RHEED patterns (Fig. S2) of the GaN template and GaN buffer layer, plots of the temporal evolution of the Ga adlayer and excess coverage predicted by the simulations (Fig. S3), and a table compiling the physically relevant simulation parameters (Table S1).

## ACKNOWLEDGMENTS

The authors thank Carsten Stemmler, Hans-Peter Schönherr, and Claudia Herrmann for MBE maintenance, Patrick Vogt for a critical reading of the manuscript, and Vladimir Kaganer for valuable discussions. Funding by the German Federal Ministry of Research, Technology, and Space (BMFTR) as well as the Berlin Senate is gratefully acknowledged.

## CONFLICT OF INTEREST

The authors have no conflicts to disclose.

# SUPPLEMENTARY MATERIAL

## Simultaneously monitoring Ga adsorption and desorption kinetics on GaN(0001) using four in situ techniques

Huaide Zhang,* Philipp John, Jingxuan Kang, Lutz Geelhaar, Yongjin Cho, and Oliver Brandt
*Paul-Drude-Institut für Festkörperelektronik, Leibniz-Institut im Forschungsverbund Berlin e.V., Hausvogteiplatz 5-7, 10117 Berlin, Germany*

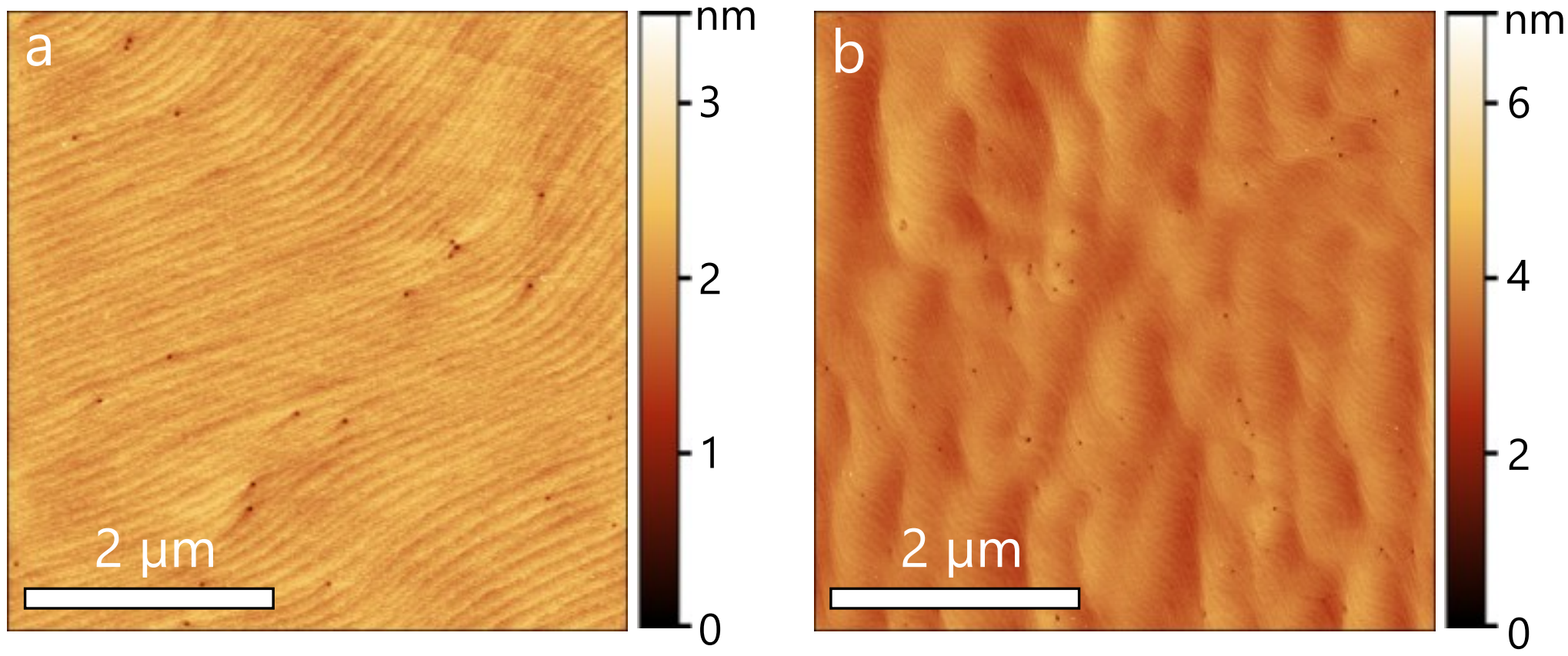

FIG. S1. **Surface morphology of GaN template and buffer layer.** Topographs of the (a) MOCVD-grown GaN template and (b) MBE-grown GaN buffer layer recorded by atomic force microscopy. The surface of both layers exhibits a morphology characteristic for step-flow growth with nanoscopic pits originating from the outcrops of threading dislocations. The root-mean-square roughness of the template and the buffer layer is 0.16 and 0.37 nm, respectively.

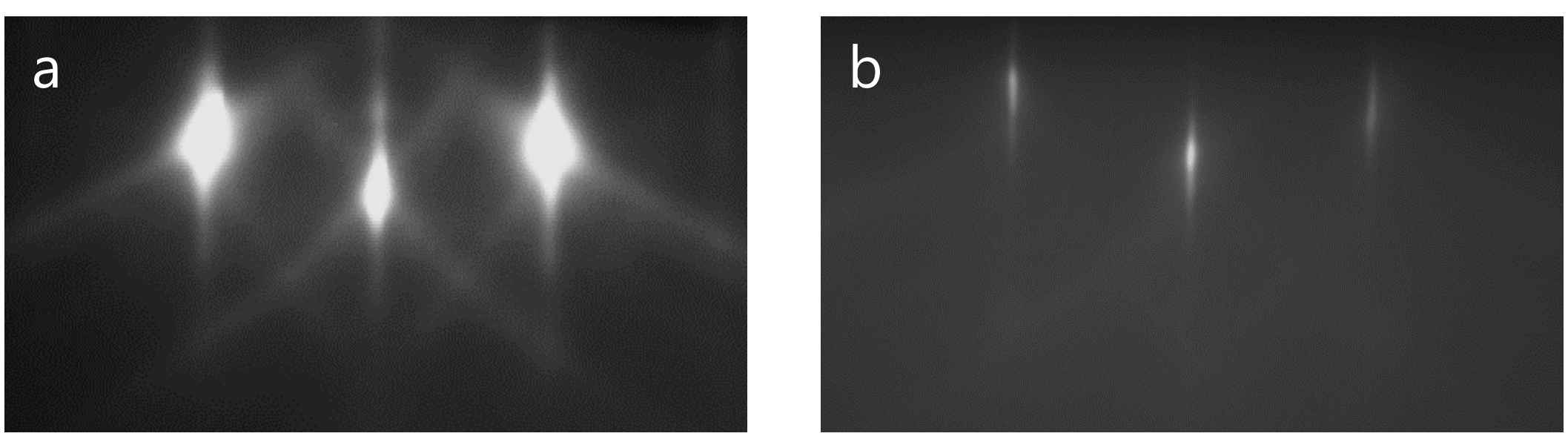

FIG. S2. **Impact of the Ga adlayer on RHEED intensity.** RHEED patterns along the $[11\bar{2}0]$ azimuth of the (a) static GaN(0001) surface without Ga adatoms and (b) the GaN(0001) surface covered with a complete Ga bilayer during Ga-stable MBE growth.

* Electronic mail: zhang@pdi-berlin.de

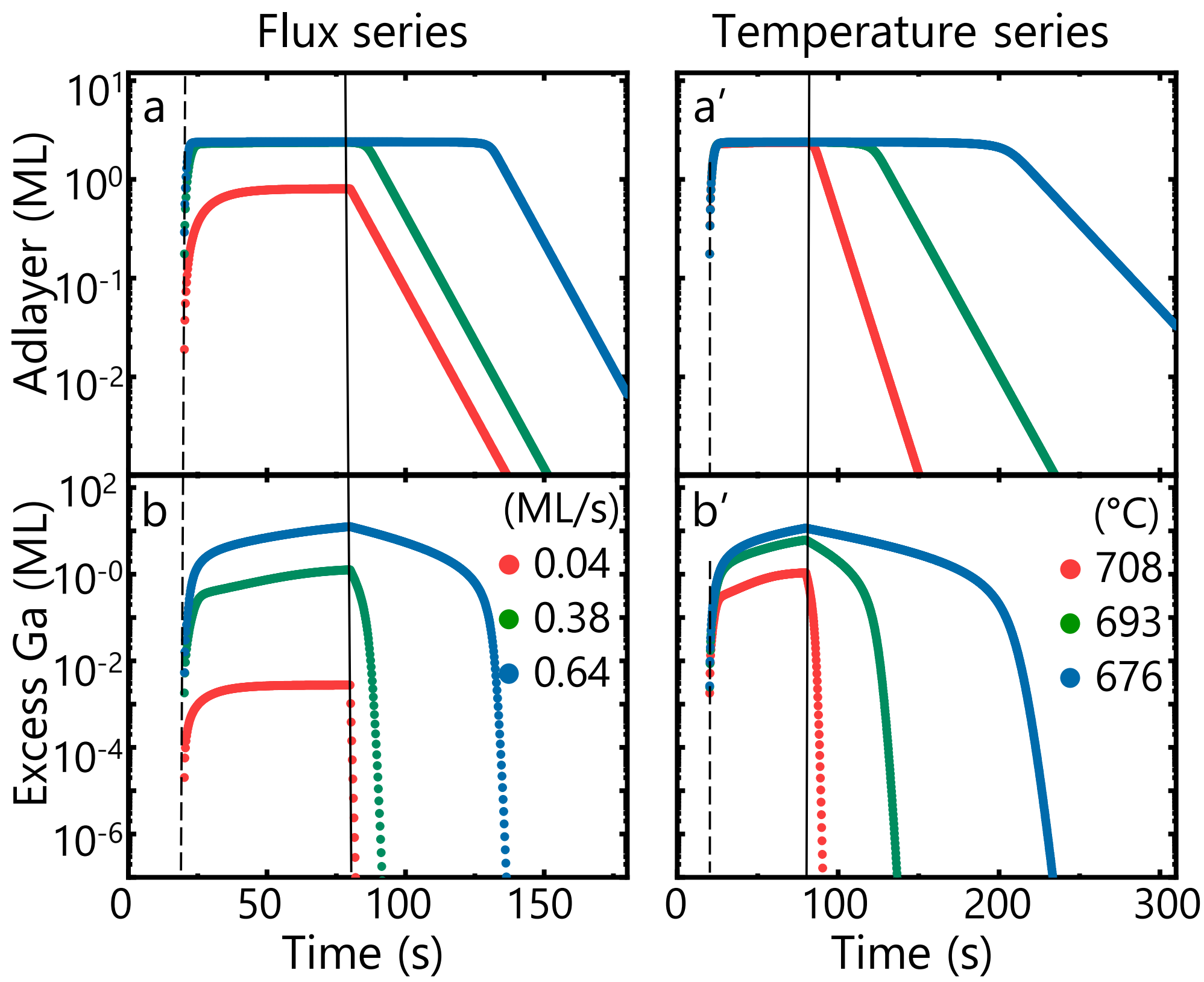

FIG. S3. **Temporal evolution of Ga coverage.** (a,a') Simulated Ga adlayer coverage in the (a) flux and (a') temperature series. (b, b') Simulated Ga coverage in excess of the adlayer in the two series. The vertical dashed and solid lines indicate the times $t$ for opening ($t$ = 20 s) and closing ($t$ = 80 s) the Ga shutter, respectively.

TABLE S1. **Physically relevant parameters for the simulations.** The rate constants $\gamma$, $\kappa_0$ and $\delta$ that were used to simulate the transients of the four *in situ* techniques in (a) the flux series and (b) the temperature series as shown in Fig. 2 of the main text.

(a) Flux series (708 °C)

| Flux (ML/s) | $\gamma$ ($s^{-1}$) | $\kappa_0$ ($s^{-1}$) | $\delta$ ($s^{-1}$) |
|---|---|---|---|
| 0.04 | 0.118 | N/A | 7 |
| 0.38 | 0.118 | 0.15 | 7 |
| 0.64 | 0.118 | 0.09 | 7 |

(b) Temperature series (0.38 ML/s)

| Temperature (°C) | $\gamma$ ($s^{-1}$) | $\kappa_0$ ($s^{-1}$) | $\delta$ ($s^{-1}$) |
|---|---|---|---|
| 708 | 0.118 | 0.15 | 7 |
| 693 | 0.068 | 0.12 | 2 |
| 676 | 0.04 | 0.042 | 1 |